\RequirePackage{lineno}
\documentclass[aps,prl,twocolumn,showpacs,showkeys,amsmath,amssymb,superscriptaddress,groupedaddress]{revtex4}  
\usepackage{graphicx}  
\usepackage{dcolumn}   
\usepackage{bm}        
\usepackage{amssymb}   
\usepackage{lineno}
\usepackage{multirow}
\newcommand{\result}{$0.083\pm0.018$}

\newcommand{\combin}{$0.089\pm0.008$}

\begin{document}
\title{Independent Measurement of the Neutrino Mixing Angle $\theta_{13}$ via Neutron Capture on Hydrogen at Daya Bay}
\newcommand{\ECUST}{\affiliation{Institute of Modern Physics, East China University of Science and Technology, Shanghai}}
\newcommand{\IHEP}{\affiliation{Institute~of~High~Energy~Physics, Beijing}}
\newcommand{\Wisconsin}{\affiliation{University~of~Wisconsin, Madison, Wisconsin, USA}}
\newcommand{\Yale}{\affiliation{Department~of~Physics, Yale~University, New~Haven, Connecticut, USA}}
\newcommand{\BNL}{\affiliation{Brookhaven~National~Laboratory, Upton, New York, USA}}
\newcommand{\NTU}{\affiliation{Department of Physics, National~Taiwan~University, Taipei}}
\newcommand{\NUU}{\affiliation{National~United~University, Miao-Li}}
\newcommand{\Dubna}{\affiliation{Joint~Institute~for~Nuclear~Research, Dubna, Moscow~Region}}
\newcommand{\CalTech}{\affiliation{California~Institute~of~Technology, Pasadena, California, USA}}
\newcommand{\CUHK}{\affiliation{Chinese~University~of~Hong~Kong, Hong~Kong}}
\newcommand{\NCTU}{\affiliation{Institute~of~Physics, National~Chiao-Tung~University, Hsinchu}}
\newcommand{\NJU}{\affiliation{Nanjing~University, Nanjing}}
\newcommand{\TsingHua}{\affiliation{Department~of~Engineering~Physics, Tsinghua~University, Beijing}}
\newcommand{\SZU}{\affiliation{Shenzhen~University, Shenzhen}}
\newcommand{\NCEPU}{\affiliation{North~China~Electric~Power~University, Beijing}}
\newcommand{\Siena}{\affiliation{Siena~College, Loudonville, New York, USA}}
\newcommand{\IIT}{\affiliation{Department of Physics, Illinois~Institute~of~Technology, Chicago, Illinois, USA}}
\newcommand{\LBNL}{\affiliation{Lawrence~Berkeley~National~Laboratory, Berkeley, California, USA}}
\newcommand{\UIUC}{\affiliation{Department of Physics, University~of~Illinois~at~Urbana-Champaign, Urbana, Illinois, USA}}
\newcommand{\CDUT}{\affiliation{Chengdu~University~of~Technology, Chengdu}}
\newcommand{\RPI}{\affiliation{Department~of~Physics, Applied~Physics, and~Astronomy, Rensselaer~Polytechnic~Institute, Troy, New~York, USA}}
\newcommand{\SJTU}{\affiliation{Shanghai~Jiao~Tong~University, Shanghai}}
\newcommand{\BNU}{\affiliation{Beijing~Normal~University, Beijing}}
\newcommand{\WM}{\affiliation{College~of~William~and~Mary, Williamsburg, Virginia, USA}}
\newcommand{\Princeton}{\affiliation{Joseph Henry Laboratories, Princeton~University, Princeton, New~Jersey, USA}}
\newcommand{\VirginiaTech}{\affiliation{Center for Neutrino Physics, Virginia~Tech, Blacksburg, Virginia, USA}}
\newcommand{\CIAE}{\affiliation{China~Institute~of~Atomic~Energy, Beijing}}
\newcommand{\SDU}{\affiliation{Shandong~University, Jinan}}
\newcommand{\NanKai}{\affiliation{School of Physics, Nankai~University, Tianjin}}
\newcommand{\UC}{\affiliation{Department of Physics, University~of~Cincinnati, Cincinnati, Ohio, USA}}
\newcommand{\DGUT}{\affiliation{Dongguan~University~of~Technology, Dongguan}}
\newcommand{\XJTU}{\affiliation{Xi'an Jiaotong University, Xi'an}}
\newcommand{\UCB}{\affiliation{Department of Physics, University~of~California, Berkeley, California, USA}}
\newcommand{\HKU}{\affiliation{Department of Physics, The~University~of~Hong~Kong, Pokfulam, Hong~Kong}}
\newcommand{\UH}{\affiliation{Department of Physics, University~of~Houston, Houston, Texas, USA}}
\newcommand{\Charles}{\affiliation{Charles~University, Faculty~of~Mathematics~and~Physics, Prague}}
\newcommand{\USTC}{\affiliation{University~of~Science~and~Technology~of~China, Hefei}}
\newcommand{\TempleUniversity}{\affiliation{Department~of~Physics, College~of~Science~and~Technology, Temple~University, Philadelphia, Pennsylvania, USA}}
\newcommand{\CUC}{\affiliation{Institute of Physics, Pontifical Catholic University of Chile, Santiago, Chile}} 
\newcommand{\CGNPG}{\affiliation{China~Guangdong~Nuclear~Power~Group, Shenzhen}}
\newcommand{\NUDT}{\affiliation{College of Electronic Science and Engineering, National University of Defense Technology, Changsha}} 
\newcommand{\IowaState}{\affiliation{Iowa~State~University, Ames, Iowa, USA}}
\newcommand{\ZSU}{\affiliation{Sun Yat-Sen (Zhongshan) University, Guangzhou}}
\author{F.~P.~An}\ECUST
\author{A.~B.~Balantekin}\Wisconsin
\author{H.~R.~Band}\Wisconsin
\author{W.~Beriguete}\BNL
\author{M.~Bishai}\BNL
\author{S.~Blyth}\NTU
\author{I.~Butorov}\Dubna
\author{G.~F.~Cao}\IHEP
\author{J.~Cao}\IHEP
\author{Y.L.~Chan}\CUHK
% ...
\collaboration{Daya Bay Collaboration}
\noaffiliation
\date{\today}

\begin{abstract}
A new measurement of the $\theta_{13}$ mixing angle has been obtained at
the Daya Bay Reactor Neutrino Experiment via the detection of
inverse beta decays tagged by neutron capture on hydrogen.
The antineutrino events for hydrogen capture are
distinct from those for gadolinium capture with largely different
systematic uncertainties, allowing a determination independent of the gadolinium-capture result and an improvement on the precision of $\theta_{13}$
measurement.
With a 217-day antineutrino data set obtained with six antineutrino detectors and from six 2.9 GW$_{th}$ reactors,
the rate deficit observed at the far hall is interpreted as
$\sin^22\theta_{13}=$\result ~in the three-flavor oscillation model.
When combined with the gadolinium-capture result from
Daya Bay, we obtain $\sin^22\theta_{13}$=\combin ~as the final result
for the six-antineutrino-detector configuration of the Daya Bay experiment.
\end{abstract}


\maketitle

Neutrino oscillations are described by the
three angles ($\theta_{13}$, $\theta_{23}$, $\theta_{12}$) and phase ($\delta$) of the PMNS matrix~\cite{pontecorvo, mns}.
Recent results~\cite{DYB1, DYB2, DYB3, RENO, T2Kn} have established that $\theta_{13}$ is non-zero, as had been
indicated by accelerator- and reactor-neutrino experiments~\cite{T2K, MINOS, MINOS2, KL, DC1, DC2, DCnH}.
Accurate and precise knowledge of $\theta_{13}$ is essential to forthcoming experiments to
determine the neutrino mass hierarchy and to search for CP violation in the lepton sector~\cite{LBNE}.
Definite $\theta_{13}$ results were obtained by measuring the
changes of reactor antineutrino rates and spectra at multiple sites
via the inverse-beta decay (IBD) reaction,
$\bar\nu_e + p \rightarrow e^+ + n$, in which the prompt $e^+$ signal
is tagged by the delayed $\sim$8 MeV $\gamma$-cascade signal from neutron
capture on gadolinium (nGd)~\cite{DYB1, DYB2, DYB3, RENO}.
In this Letter, with comparable statistics as the nGd case,
a new measurement obtained by tagging the delayed 2.2 MeV $\gamma$
from neutron capture on hydrogen (nH)~\cite{Kamland, SuperK, DCnH}
at Daya Bay is presented.
New analysis approaches have been developed to
meet the challenges associated with the higher background, longer neutron
capture time ($\sim$200 $\mu$s), and a lower energy $\gamma$ ray
from neutron capture for nH IBD events.
This nH analysis provides an independent measurement of
$\sin^22\theta_{13}$, and leads to an improved precision on the $\theta_{13}$
mixing angle when combined with the nGd result obtained from
the same period of the six antineutrino detector (AD) configuration~\cite{DYB3}.
The inclusion of nH capture results will improve the ultimate precision of Daya Bay for both $\theta_{13}$ and the $\bar\nu_e$ mass-squared difference $|\Delta m_{ee}^2|$~\cite{DYB3}.
Optimization of the nH analysis method will be applicable to future
reactor neutrino experiments that address the reactor-antineutrino anomaly~\cite{anomally, anomally2, anomally3, anomally4}
and determine the neutrino mass hierarchy~\cite{hierarchy, hierarchy2, hierarchy3, hierarchy4}.

A detailed description of the Daya Bay experiment can be found in~\cite{DNIM, TDR}.
The ongoing experiment consists of two near experimental halls, EH1 and EH2,
and one far hall, EH3.
The power-weighted baselines to the six commercial power reactors are
$\sim$500~m and $\sim$1.6~km for the near and far halls, respectively.
In this analysis, EH1, EH2 and EH3 have two, one and three ADs, respectively.
All ADs are submerged
in water pools consisting of optically separated inner (IWS) and outer
water shields (OWS), which also function as Cherenkov detectors to
tag cosmic-ray muons.
All ADs utilize an identical three-zone design with 20 tons of Gd-loaded
liquid scintillator (GdLS) in the innermost zone,
22 tons of liquid scintillator (LS) in the middle zone
to detect $\gamma$'s escaped from GdLS,
and 40 tons of mineral oil in the outermost zone where photo-multiplier
tubes (PMTs) are installed.
Unlike the nGd events, nH capture can occur both in the LS and
the GdLS regions,
resulting in more nH than nGd events before event selection.
The trigger threshold for each AD was set at $\sim$0.4 MeV based
on the logical OR of the number of over-threshold PMTs
and the analog sum of their signals~\cite{trg}.
The vertex and energy were reconstructed utilizing
the charge topological information collected by the PMTs.
For a 2.2-MeV $\gamma$,
the vertex resolutions were $\sim$8~cm in the $x$-$y$ plane and $\sim$13~cm in the $z$ direction
in a Cartesian coordinate system
with the origin at the AD center and the $+z$ axis pointing upwards.
Detector simulation was based on GEANT4~\cite{G4} with
the relevant physical processes validated~\cite{DNIM}.
All data from Dec.\ 24, 2011 to Jul.\ 28, 2012 were used
for this analysis.
The live time of each AD is listed in Table~\ref{tab:HSummary}.

All triggered events at each site were
sequenced according to their time stamps
after removing an instrumental background resulting from spontaneous light emission of
PMTs~\cite{DYB1, DYB2}.
Because of the latency between detectors, events with time separations less than 2 $\mu$s in the same hall were grouped together for identifying cosmic-ray muons.
A water-pool muon was defined as an event with the number of
over-threshold PMTs $>$12 in the IWS or $>$15
in the OWS, while an AD (shower) muon had a visible energy greater
than 20 MeV (2.5 GeV) in an AD.
Table~\ref{tab:HSummary} lists the total muon rate per AD, $R_\mu$,
which was stable over the entire data-taking period.
Due to the long lifetimes of muon spallation products,
the AD events were required to occur at least 400~$\mu$s, 800~$\mu$s or 1~s after a water-pool, AD or
shower muon, respectively.
The visible energy for each AD event was also
required to be greater than 1.5 MeV to reject the low-energy background.
The surviving AD events were denoted as ``good''
events for further study.
Coincident events were identified within a 399 $\mu$s time window, $T_c$,
beginning at 1 $\mu$s after each prompt signal candidate~\cite{acc}.
This procedure classified all good events into single-coincidence,
double-coincidence (DC), and multi-coincidence categories.
Events in the latter category account for $\sim$2\% of the total and
were not included for further analysis.

\begin{figure}[!t]
\includegraphics[width=\columnwidth]{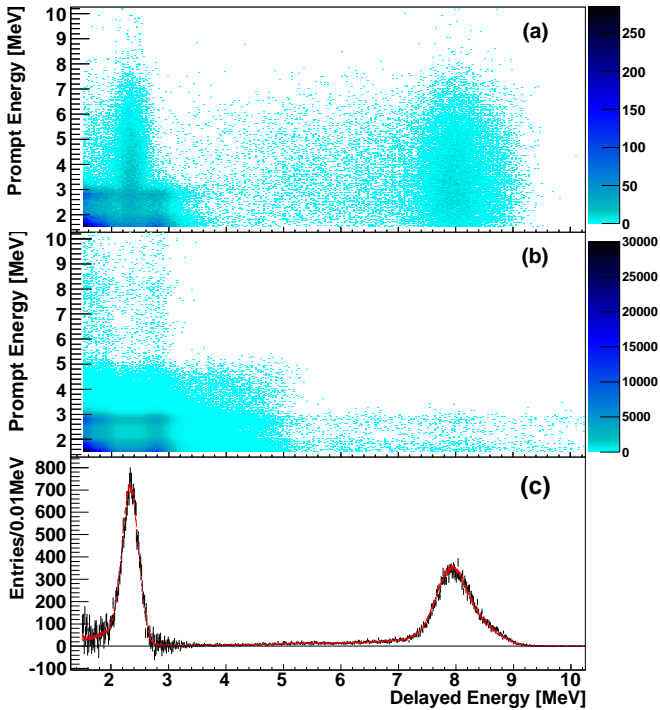}
\caption{(color online) (a) The prompt {\it vs.}\ delayed energy of double coincidence events with a maximum 50 cm vertex separation for all far-hall ADs,
(b) the accidental background sample (ABS) events and (c) the delayed energy distribution after subtracting the accidentally coincident background
for the far hall (black) and the near halls (red), where the total near site
spectrum was normalized to the area of the far site spectrum.}
\label{fig:NF}
\end{figure}

Since the DC events were dominantly accidentally coincident background,
especially in the far hall,
a maximum distance of 50 cm between the prompt and delayed vertices
was required,
rejecting 98\% of this background at the cost of a 25\% acceptance loss.
This cut was one of the major differences between
the nH and the nGd analyses.
Figure~\ref{fig:NF} (a) shows the prompt energy {\it vs.}\ the delayed energy
for all the DC events after this cut in the far hall.
The IBD bands are clearly seen for both the 2.2-MeV-nH and the
8-MeV-nGd cases.
The measured nH peak was around 2.33 MeV with a resolution
of 0.14 MeV.
The offset from the true peak value arose from
the nonlinear and nonuniform energy response, which was pegged to the nGd capture peak in the reconstruction.
The $\gamma$'s from $^{40}$K and $^{208}$Tl decays are observed
around 1.5 and 2.6 MeV, respectively,
and the continuous bands from 1.5 to 3 MeV are from the decay
products of $^{238}$U and $^{232}$Th.
The nH IBD candidates were obtained by requiring the prompt
energy to be less than 12 MeV and
the delayed energy to be within $\pm3\sigma$ of the measured
nH peak in each AD.
The numbers of the candidates are listed in Table~\ref{tab:HSummary}.

The four identified backgrounds in the selected sample are
accidental coincidences, cosmogenically produced fast neutrons and
$^9$Li/$^8$He, and neutrons from the retracted
$^{241}$Am-$^{13}$C calibration source.
The delayed signals of the latter three are all from correlated neutron captures.

The following procedure was adopted for removing the accidental
coincidence background.
An accidental background sample (ABS) consisting of $N_{\rm ABS-tot}$
events was first generated by pairing
two single events separated by at least 10 hours.
The same distance and energy cuts were then applied to the ABS events,
resulting in $N_{\rm ABS-cut}$ events.
As shown in Fig.~\ref{fig:NF} (b), the ABS describes well the
pattern of the low-energy region in Fig.~\ref{fig:NF} (a).
The spectra of correlated events dominated by IBD, $N_{\rm IBD}(\xi)$, were
then obtained by subtracting the accidental background
from the DC events, $N_{\rm DC}$:

\begin{equation}
N_{\rm IBD}(\xi)=N_{\rm DC}(\xi) - R \cdot T_{\rm live}
\cdot\frac{N_{\rm ABS-cut}(\xi)}{N_{\rm ABS-tot}},
\label{equ:sub}
\end{equation}
where $\xi$ represents the quantity under study,
such as the delayed energy,
$T_{\rm live}$ is the live time of
data-taking listed in Table~\ref{tab:HSummary}, and $R$ is the
random coincidence rate
that can be written as~\cite{acc}
\begin{align}
\label{equ:Acc}
R=R_s \times e^{-R_s T_c} \times R_sT_c e^{-R_sT_c},
\end{align}
where $R_s$ is the singles rate,
$e^{-R_sT_c}$ gives the probability of no prior coincidence within $T_c$,
and $R_sT_c e^{-R_sT_c}$ is the probability of a trigger from an accidental
coincidence within $T_c$.
Table~\ref{tab:HSummary} lists the average rate of the
accidental background in eq.~(\ref{equ:Acc}) for each AD.

While the statistical uncertainty of $R_s$ is negligible,
a systematic uncertainty is caused by the presence in the
single event sample of
a very small fraction of genuine correlated events for which
either the prompt or the delayed event is not detected.
The singles rate $R_s$ was determined to be $\sim$22 Hz from the average of
the good triggered event rates before and after excluding both the
DC events and the multi-coincidence events.
The systematic uncertainty in $R_s$, estimated
from the difference of these two rates, was found to be
0.18\%, 0.16\% and 0.05\% for the EH1, EH2 and EH3, respectively.
The singles rate $R_s$ was observed to have a slow downward trend ($<0.36$\%/day) immediately
after an AD was installed in water and become stable after about four months.
The slow variation of $R_s$ was taken into account by performing the
accidental subtraction, eq.~(\ref{equ:sub}), on a run-by-run basis,
with each run lasting about two days.

\begin{figure}[!t]
\includegraphics[angle=270, width=\columnwidth]{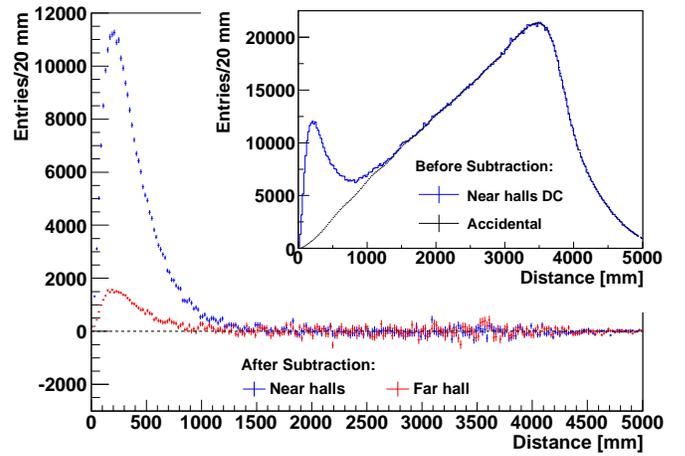}
\caption{(color online) Distributions of the distance between the prompt and the delayed vertices after the accidental background was subtracted for the near
halls (blue) and the far hall (red). The inset plot shows the distance distributions for both the near halls double coincidence, DC, events (blue)
and the expected accidental background sample (black).}
\label{fig:distance}
\end{figure}

Figure~\ref{fig:NF} (c) shows the delayed energy spectra for the DC events
in the near and far halls after subtracting the accidental background.
Very similar spectra, clearly showing the nH and nGd peaks, were
observed for all ADs.
The procedure of accidental background subtraction was validated
by checking the distribution of distance between the prompt and delayed
vertices as shown in Fig.~\ref{fig:distance}.
Simulation studies indicated IBD events rarely occurred with the
prompt and delay vertices separated beyond $200$ cm.
Figure~\ref{fig:distance} shows
a flat distribution consistent with zero for the region beyond $200$ cm.
The subtraction procedure was further validated from the distribution of neutron capture time.
The accidental-background-subtracted spectra are consistent with no
events of coincidence time longer than 1.5~ms.

The procedures for evaluating the $^9$Li/$^8$He, fast neutron,
and $^{241}$Am-$^{13}$C backgrounds follow those in~\cite{DYB1}, except for
three different selection cuts: the delayed energy cut,
the distance cut,
and an additional cut, $E>3.5$~MeV, on the prompt energy to suppress
the accidental background.
The fast-neutron background is significantly higher than in the nGd case because the
LS region is
more accessible to the externally produced fast neutrons.
The other two backgrounds are also slightly different due to
detector geometry configuration.
All background
rates are listed in Table~\ref{tab:HSummary}.

\begin{table*}[htb]
\center
\begin{tabular}{l c c c c c c }\hline\hline
 & \multicolumn{2}{c}{EH1} &  EH2 & \multicolumn{3}{c}{EH3} \\
 & AD1 & AD2 & AD3 & AD4 & AD5 & AD6 \\\hline
Live time (day) & 191.0 & 191.0 & 189.6 & 189.8 & 189.8 & 189.8 \\
$R_{\mu}$ (Hz) & 201.0 & 201.0 & 150.6 & 15.73 & 15.73 & 15.73 \\
$\varepsilon_{\mu}\varepsilon_{m}$  & 0.7816 & 0.7783 & 0.8206 & 0.9651 & 0.9646 & 0.9642 \\
Candidates & 74136 & 74783 & 69083 & 20218 & 20366 & 21527 \\
Accidental rate (/AD/day)& $64.96\pm0.13$ & $64.06\pm0.13$ & $57.62\pm0.11$ & $62.10\pm0.06$ & $64.05\pm0.06$ & $68.20\pm0.07$ \\
Fast n rate (/AD/day) & \multicolumn{2}{c}{$2.09\pm0.56$} & $1.37\pm0.40$ & \multicolumn{3}{c}{$0.10\pm0.04$} \\
$^9$Li/$^8$He rate (/AD/day) & \multicolumn{2}{c}{$2.75\pm1.38$} &  $2.14\pm1.07$ & \multicolumn{3}{c}{$0.26\pm0.13$} \\
$^{241}$Am-$^{13}$C rate (/AD/day) & $0.09\pm0.05$ & $0.09\pm0.05$ & $0.09\pm0.05$ & $0.06\pm0.03$ & $0.06\pm0.03$ & $0.06\pm0.03$ \\
IBD rate (/AD/day) & $426.71\pm2.36$ & $434.09\pm2.37$ & $382.69\pm2.04$ & $47.87\pm0.79$ & $46.78\pm0.79$ & $49.02\pm0.82$ \\
nH/nGd  & $0.653\pm0.004$ & $0.654\pm0.004$ & $0.658\pm0.004$ & $0.653\pm0.012$ & $0.641\pm0.012$ & $0.679\pm0.013$ \\\hline
\end{tabular}
\caption{Summary of the hydrogen capture data sample. All the rate quantities are corrected with $\varepsilon_{\mu}\varepsilon_{m}$.
The bottom row contains the ratio of the measured nH IBD rate to that of nGd from~\cite{DYB3}.}
\label{tab:HSummary}
\end{table*}


The number of
predicted IBD events, $N$, summed over various detector
volumes $v$ (GdLS, LS, and acrylic vessels)
is given as

\begin{align}
\label{equ:effi}
    N
    &= \phi \sigma \varepsilon_{\mu} \varepsilon_{m} \left[ \sum_v^{\text{GdLS, LS, Acry.}} N_{p,v} f_v \varepsilon_{ep,v}
\varepsilon_{ed,v} \varepsilon_{t,v} \right] \varepsilon_{d},
\end{align}
where $\phi$ is the antineutrino flux, which was modeled as in~\cite{DYB3},
and $N_p$, $\sigma$ and $f$ are the number of protons, IBD cross section
and hydrogen capture fraction, respectively.
The efficiency $\varepsilon_{\mu}$ is the efficiency of the muon veto
and $\varepsilon_{m}$ is the efficiency
of the multiplicity cut for the DC selection~\cite{acc}.
The efficiency $\varepsilon_{ep}$ ($\varepsilon_{ed}$) is the prompt (delayed)
energy cut efficiency, and $\varepsilon_{t}$ ($\varepsilon_{d}$) refers to
the efficiency of the time (distance) cut.

The $\theta_{13}$ analysis is based on relative rates,
as in~\cite{DYB1, DYB2}, such that
uncertainties that are correlated among ADs largely cancel and
the uncorrelated uncertainties give the dominant contributions.

The central values of $\varepsilon_{ep}$ and $\varepsilon_{ed}$ were
evaluated from the simulation. The prompt energy cut at 1.5 MeV
caused about 5\% inefficiency in $\varepsilon_{ep}$ for GdLS and LS events
and a much higher loss in the acrylic.
The slight variations in energy scale and resolution among different
ADs introduced an uncorrelated uncertainty of 0.1\%.
For $\varepsilon_{ed}$, the 3-$\sigma$ energy cut around the nH
capture peak made the efficiency largely insensitive to the
small variations of energy calibration and resolution.
The efficiency $\varepsilon_{ed}$ also included a small contribution from the
low energy tail of nGd capture events.
The uncertainty in $\varepsilon_{ed}$ was determined by
using a spallation neutron sample.
Since the spallation neutron fluxes for neighboring ADs were
nearly identical
and the relative nGd acceptance in the GdLS region was accurately
measured~\cite{DYB1, DYB2},
a comparison of the spallation neutron rates between nH and nGd
captures gave an uncertainty of 0.5\%.
Simulations of IBD events in different ADs
with as-built dimensions were also consistent with
this uncertainty estimate.

The central value of $\varepsilon_{t}$ was also evaluated with the simulation.
The sources of the uncorrelated uncertainty include the number densities of
various isotopes in LS and GdLS,
the neutron elastic and capture cross sections, and the precision
of time measurements.
A chemical analysis showed that the density difference among the
ADs is less than 0.1\% and
that the weight fractions of carbon and hydrogen among the ADs
differed by less than 0.3\%, limited by the instrumental precision.
The uncertainty in number densities introduced a 0.1\%
uncorrelated uncertainty in $\varepsilon_{t}$.
The precision of the timing measurement was studied using
$\beta$-$\alpha$ coincident events from the decay chain
of $^{214}$Bi-$^{214}$Po-$^{210}$Pb
originating from the $^{238}$U cascade decays.
With the same procedure of accidental subtraction applied, a comparison of
the measured lifetime of $^{214}$Po with the known value (237 $\mu$s)
verified that the uncertainty on the timing precision due to the
electronics was at the level of 0.1\%.
In total, the uncorrelated uncertainty was taken as 0.14\%.
A study of a clean nH IBD sample with the prompt energy $>$3.5~MeV
for the ADs in the two near halls also confirmed this conclusion.

The central value of $\varepsilon_{d}$ was directly measured from the
distribution of the distance
between the prompt and delayed vertices (see Fig.~\ref{fig:distance}).
The uncorrelated uncertainty, caused by the slight variations in the
vertex reconstruction bias and resolution, was estimated to be 0.4\%.


The value and uncertainty of $N_p$ in GdLS were discussed in~\cite{DNIM}.
The proton number $N_p$ in the LS region was determined in the same way and
its uncorrelated uncertainty of 0.13\% was dominated
by the uncertainty of the Coriolis-mass-flow meter.
The H-capture fraction, $f$, was less than unity due to neutron capture on Gd and C,
and was estimated by the simulation
to be 96\% in the LS region and 16\% in the GdLS region.
The relative difference among ADs is negligible \cite{DYB2}.

The selected nH IBD sample was about 65\% of the size of the nGd IBD sample~\cite{DYB3}.
The total uncorrelated uncertainty per AD was 0.67\% as summarized in Table~\ref{tab:HEff}.
The nH/nGd ratios among ADs 1, 2, and 3 agreed within 0.6\% as shown in Table~\ref{tab:HSummary}, which
provided a strong confirmation of the uncorrelated uncertainty per AD.

\begin{table}[h]
\begin{tabular}{ l c  c  c  }\hline\hline
\multicolumn{1}{r}{}                 & Uncorrelated uncertainty  & Coupled\\\hline
$N_{p,\rm {GdLS}}$     & 0.03\%  & yes \\
$N_{p,\rm {LS}}$       & 0.13\%  & no \\
$N_{p,\rm {Acrylic}}$  & 0.50\%  & no \\\hline
{$\varepsilon_{ep,v}$} & 0.1\%   & yes \\
{$\varepsilon_{ed,v}$} & 0.5\%   & no  \\
{$\varepsilon_{t,v}$}  & 0.14\%  & yes \\
$\varepsilon_{d}$      &  0.4\%  & no \\\hline
Combined               & 0.67\%        &    \\\hline\hline
\end{tabular}
\caption{
The per-AD relative uncorrelated uncertainty summary. The quoted uncertainties
on the efficiencies are independent of volume. The combined uncertainty
takes into account the relative GdLS, LS and acrylic masses.
The last column indicates whether the uncorrelated uncertainties for the nH and nGd analyses are
coupled.}
\label{tab:HEff}
\end{table}

\begin{figure}[!ht]
\includegraphics[width=\columnwidth, angle=270]{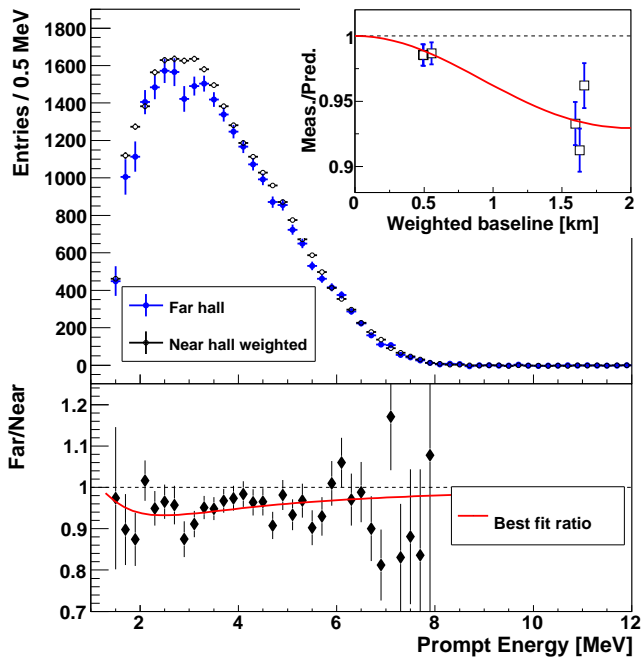}
\caption{(color online) The detected energy spectrum of the prompt events of the far hall ADs (blue) and
near hall ADs (open circle) weighted according to baseline.
The far-to-near ratio (solid dot) with best fit $\theta_{13}$ value is shown in the lower plot.
In the inset is the ratio of the measured to the predicted rates in
each AD {\it vs.}\ baseline, in which the AD4 (AD6) baseline was shifted relative to that of AD 5 by 30 ($-30$) m. }
\label{fig:FNRatio}
\end{figure}

Figure~\ref{fig:FNRatio} shows a comparison of the prompt spectra of the far hall
and the near halls weighted by the near-to-far baseline ratio,
along with the ratio of the measured-to-predicted rates as a function of baseline.
Clear evidence for electron antineutrino disappearance is observed.
A $\chi^2$ with pull terms for nuisance parameters
as in~\cite{DYB1, DYB2} is minimized to extract $\sin^22\theta_{13}$ from the
detected nH IBD rate deficit.
The value of $|\Delta m^2_{31}|$ is taken from MINOS~\cite{dm31}.
The best fit is $\sin^22\theta_{13}$=\result ~with $\chi^2$=4.5 for 4
degrees of freedom.
The increase in $\chi^2$ is 20 when $\theta_{13}$ is set to zero,
ruling out this null assumption at 4.6 standard deviations.
The expected Far/Near ratio based on the best-fit $\sin^22\theta_{13}$ value
is compared to data in Fig.~\ref{fig:FNRatio}.

\vspace*{0.5cm}

The nH result is an independent measurement of $\theta_{13}$ and
provides a strong confirmation of the earlier measurement using nGd~\cite{DYB3}.
Currently both the nH and nGd~\cite{DYB3} uncertainties are statistics-dominated.
With only statistical uncertainties considered in the nH fit, the
uncertainty of $\sin^22\theta_{13}$ is 0.015,
about 70\% of the total uncertainty when uncertainties are added in quadrature,
which is the same for the nGd analysis.
The dominant systematic uncertainties are also independent of the nGd analysis.
For example, the delayed-energy cut is uncoupled (uncorrelated) because
the impact of the relative energy-scale difference on the fixed-energy threshold
in the nGd analysis~\cite{DYB1, DYB2, DYB3} is avoided
with the data-driven 3-$\sigma$ cut.
Further couplings are noted in the Table~\ref{tab:HEff}.
With all uncoupled uncertainties included in the nH fit,
the uncertainty of
$\sin^22\theta_{13}$ is 0.017 (90\% of the total uncertainty in quadrature).
By conservatively taking all coupled quantities to be fully coupled, the correlation
coefficient is about 0.05, indicating an essentially independent measurement of $\theta_{13}$.
The weighted average of nH and nGd~\cite{DYB3} results is \combin, improving the nGd result
precision by about 8\%.

In summary, with an nH sample obtained in the six-AD configuration, by comparing the rates of the reactor antineutrinos
at the far and near halls at Daya Bay, we report an independent measurement of sin$^22\theta_{13}$ which is in good
agreement with the one extracted from the minimally correlated nGd sample. By combining the results of the nH and
nGd samples, the precision of sin$^22\theta_{13}$ is improved. In general, with different systematic issues, results
derived from nH samples
will be important when the nGd systematic uncertainty becomes dominant in the future.
It is also expected that nH analysis will enable other neutrino measurements~\cite{anomally, hierarchy}.


Daya Bay is supported in part by the Ministry of Science
and Technology of China, the United States Department of
Energy, the Chinese Academy of Sciences, the National Natural Science Foundation of China,
the Guangdong provincial government, the Shenzhen municipal government, the
China Guangdong Nuclear Power Group,
Key Laboratory of Particle \& Radiation Imaging (Tsinghua University), Ministry of Education,
Key Laboratory of Particle Physics and Particle Irradiation (Shandong University), Ministry of Education,
Shanghai Laboratory for Particle Physics and Cosmology, the Research Grants
Council of the Hong Kong Special Administrative Region of
China, University Development Fund of The University of
Hong Kong, the MOE program for Research of Excellence at
National Taiwan University, National Chiao-Tung University,
and NSC fund support from Taiwan, the U.S. National Science Foundation, the Alfred P. Sloan Foundation, the Ministry
of Education, Youth and Sports of the Czech Republic, the Joint Institute of Nuclear Research in Dubna, Russia, the CNFC-RFBR joint research program,
National Commission of Scientific and Technological Research of Chile,
and Tsinghua University Initiative Scientific Research Program. We acknowledge Yellow River Engineering Consulting Co., Ltd. and China railway 15th Bureau Group Co.,
Ltd. for building the underground laboratory. We are grateful for the ongoing cooperation from the China Guangdong
Nuclear Power Group and China Light \& Power Company.

\end{document}